\documentclass{WileyMSP-template}
\usepackage{amssymb,amsthm,amsmath,amsfonts,physics,algorithm2e,float}
\usepackage{cite}
\usepackage{multicol}
\usepackage{float}
\usepackage{xcolor}
\usepackage{ragged2e}
\DeclareUnicodeCharacter{0301}{*************************************}
\DeclareUnicodeCharacter{0308}{\"{}}
\usepackage{ulem}

\begin{document}

\pagestyle{fancy}

\title{Picosecond-resolved entanglement distribution over an urban free-space channel}

\maketitle

\author{Alessandro Laneve$^{*,\dagger,1,2,3}$,}
\author{Fabrizio Cienzo$^{\dagger,1}$,}
\author{Santiago Gómez$^1$,}
\author{Paolo Barigelli$^1$,}
\author{Philip Menz$^4$,}
\author{Mattia Beccaceci$^1$,}
\author{Giuseppe Ronco$^1$,}
\author{Giorgia Grossi$^1$,}
\author{Ievgen Brytavskyi$^5$,}
\author{Thomas Oberleitner$^5$,}
\author{Markus Wiener$^5$,}
\author{Christian Weidinger$^5$,}
\author{Tobias M. Krieger$^5$,}
\author{Ailton Garcia Jr.$^5$,}
\author{Saimon Filipe Covre da Silva$^6$}
\author{Michele B. Rota$^1$,}
\author{Nicolò Spagnolo$^1$,}
\author{Henning Weier$^4$,}
\author{Fabio Sciarrino$^1$,}
\author{Armando Rastelli$^5$,}
\author{Rinaldo Trotta$^\ddagger$$^1$}

\begin{affiliations}
$^1$Dipartimento di Fisica, Sapienza Università di Roma, Piazzale Aldo Moro 5, 00185 Roma, Italy\\
$^2$University of Vienna, Faculty of Physics, Vienna Center for Quantum Science and Technology (VCQ), 1090 Vienna, Austria\\
$^3$Christian Doppler Laboratory for Photonic Quantum Computer, University of Vienna, Faculty of Physics, 1090 Vienna, Austria\\
$^4$Quantum Space Systems GmbH, Baierbrunner Strasse 3, 81379 Munich, Germany\\
$^5$Institute of Semiconductor and Solid State Physics, Johannes Kepler University Linz, Altenberger Straße 69, 4040 Linz, Austria\\
$^6$Instituto de Física Gleb Wataghin, Universidade Estadual de Campinas (UNICAMP), 13083-859 Campinas, Brazil\\
$^*$alessandro.laneve@univie.ac.at
$^\ddagger$ rinaldo.trotta@uniroma1.it
$^\dagger$ These authors contributed equally.
\end{affiliations}


\justifying

\begin{abstract}
Time-evolving entangled states describe quantum particles whose correlations evolve in time according to a well-defined dynamics. Such states can be generated in a variety of physical systems and are promising resources for several quantum technologies, ranging from quantum clock synchronization to quantum communication. However, their full potential is currently limited by the fact that the entanglement dynamics often occur on timescales comparable to the achievable synchronization precision, especially in experiments aimed at distributing entanglement through noisy urban channels. In this context, accurate timing is not merely a technical detail, but a fundamental requirement for faithfully observing and exploiting the underlying quantum correlations.
Here, we demonstrate the faithful distribution of a fast-evolving entangled state over a 270 m free-space channel connecting two buildings in the center of Rome. The developed system incorporates a synchronization device capable of achieving sub-50 ps timing accuracy between the two ends of the link while simultaneously supporting channel stabilization. Our results demonstrate that time-evolving entanglement can be reliably transmitted through a noisy urban free-space channel, representing an important benchmark toward long-distance free-space quantum communication and the future implementation of time-correlated entangled states in demanding scenarios such as satellite-based quantum networks.
\end{abstract}
\begin{multicols}{2}
\section{Introduction}
\begin{figure*}[!t]
\centering
\includegraphics[width=\textwidth]{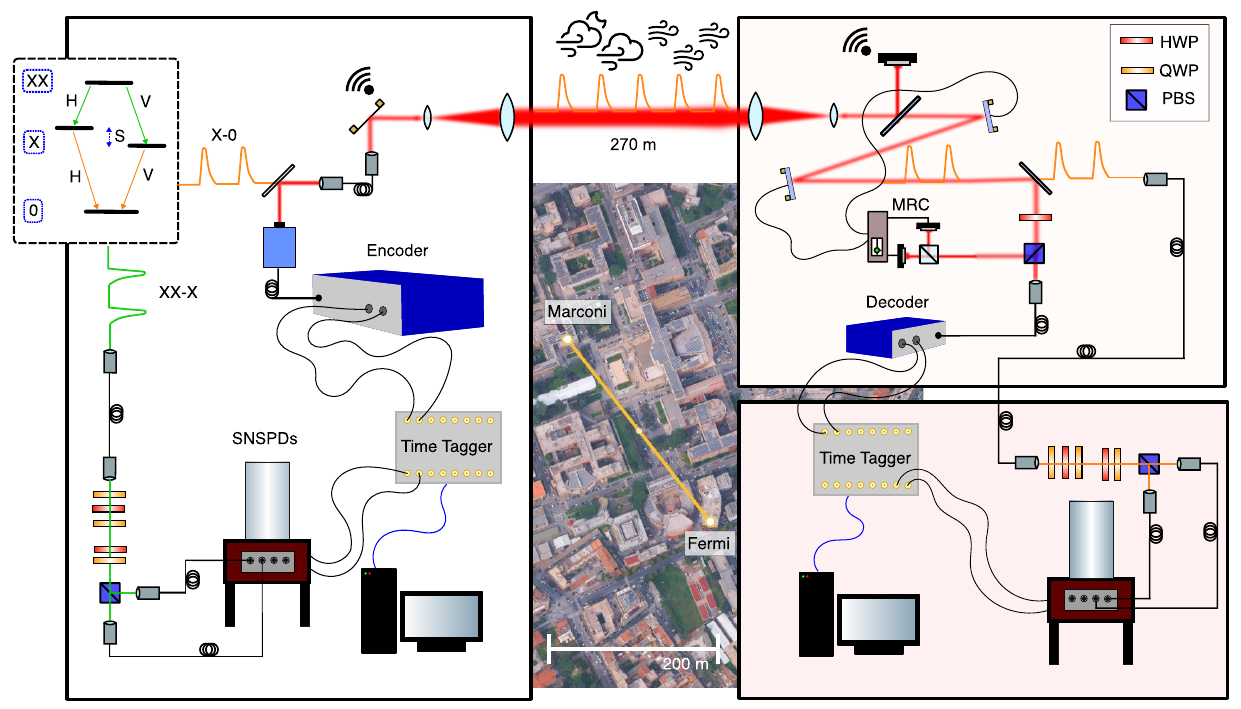}
\caption{\textbf{Scheme of the entanglement distribution setup:} \textit{the two photons emitted by the XX-X-0 cascade in a QD are spectrally separated by optical notch filters: the XX-X photons are collected by a single-mode fiber and undergo a polarization measurement, implemented by a half-waveplate (HWP), a quarter-waveplate (QWP) and a PBS. The two outputs of the PBS are collected again by single mode fibers connected to SNSPDs. The detection events are then recorded by a high time resolution correlator (Swabian TimeTagger X). The X-0 photons are sent from the Marconi to the Fermi building through the Sapienza free-space channel, after being coupled into a single mode fiber together with the 850 nm laser serving as a stabilization and synchronization beacon. 
After traveling through the 270 m channel, part of the sync laser is extracted by a dichroic mirror and directed to CCD camera which drives the slow stabilization system in the Marconi building. The rest of the sync signal is spectrally separated from the photons and sent to both a fast stabilization system (MRC Systems) and to the synchronization decoder through a multimode fiber. The splitting ratio between the signal sent to the MRC and the one sent to the decoder is controlled by a sequence of a HWP and a PBS, acting as a tunable beam-splitter. Single photons are coupled into a single mode fiber and sent to a polarization measurement station identical to the one in the Marconi building. 
The decoder device extrapolates the synchronization signal and sends it to the Fermi building correlator (Swabian TimeTagger Ultra), allowing for accurate synchronization of the detection events collected by the remote electronics, which are connected to the same Internet network and run in parallel through a server-client system. An additional QWP-HWP-QWP set is employed in both polarization measurement setups for polarization compensation. Satellite image from Google Earth\textcopyright.}}
\label{fig:figure_1_setup}
\end{figure*}
The reliable distribution of quantum entanglement over the nodes of quantum networks represents an unavoidable step towards the definitive scale up of quantum technologies to real-life scenarios, enabling distributed quantum computing \cite{cirac1999distributed}, dense coding \cite{mattle1996dense}, and fundamentally secure communication \cite{ekert1991quantum,acin2006bell}, potentially even at large distances \cite{azuma2023quantum}.  So far, many efforts have been devoted to the efficient distribution of entanglement among remote nodes, even over noisy channels and urban networks \cite{yin2012quantum,resch2005distributing,fedrizzi2009high,steinlechner2017distribution,peng2005experimental}.
For several of these applications, a common time frame is often required, as, for instance, to synchronize the measurements of communicating parties in cryptographic protocols based on the detection of quantum-correlated events.
In these protocols, the entangled state shared by the parties usually has no intrinsic temporal evolution, and synchronization is needed to effectively compare the results of the measurements. Although this task still requires the development of effective synchronization strategies, a natural question arises: is it possible to distribute a time-evolving entangled state, i.e.,  a state in which the degree of entanglement remains constant but the correlations between measurement outcomes of different observables evolve in time according to a well-defined dynamics? This type of state can be in principle engineered by combining standard entangled photon sources with fast modulation techniques in an interferometric architecture \cite{kim2017two,ecker2022remotely,xavier2025energy}, albeit with technical limitations. Alternatively, there exist quantum emitters that naturally provide time-evolving two-photon states, such as atoms \cite{aspect1984quantum, gulati2015polarization} or quantum dots (QDs) \cite{hudson2007coherence}. QDs have already been demonstrated to be suitable entangled photon sources to implement many quantum communication tasks, also in urban communication scenarios  \cite{rota2020entanglement,schimpf2021crypto,basso2021quantum,basset2023daylight,laneve2025quantum}. \\Pioneering experimental efforts \cite{pennacchietti2024oscillating, alqedra2026entanglement} and theoretical proposals \cite{shi2022clock} have indeed started to explore the possibility to employ time evolving entangled state for applications in quantum key distribution and entanglement-based clock synchronization.
However, these attempts have highlighted the crucial need for low-time jitter synchronization systems to ensure that the entanglement dynamics are accurately resolved, and thus exploited. Although many methods and devices have been developed for the synchronization of distant nodes, both in fiber-based \cite{valivarthi2022picosecond,burenkov2023synchronization},
and free-space \cite{giorgetta2013optical} practical links, the integration of such systems into entanglement distribution channels is still challenging. Deployed quantum channels often require additional ancillary devices to ensure their functioning: in free-space channels, for example, active stabilization is required \cite{fedrizzi2009high,basso2021quantum}, while adaptive phase compensation systems must be introduced in deployed fiber links \cite{neumann2022continuous,craddock2024automated}.
Despite accurate synchronization strategies have been developed in fiber-based networks \cite{alshowkan2022advanced}, low time jitter synchronization systems are still missing for free-space quantum channels, although this would represent a significant contribution towards the realization of efficient satellite-based long-range quantum networks.\\
In this work, we move a significant step in this direction, presenting a new synchronization system based on one-way time-frequency transfer, integrated in a 270 m free-space quantum link laid over the Sapienza University campus, in the city center of Rome. This system grants a 50 ps time jitter between measurement systems at the two ends of the channel, while providing at the same time a reliable beacon for stability of the link itself. We harness this system to efficiently distribute across the free-space channel an oscillating polarization entangled state generated by a semiconductor QD, as depicted in Fig. \ref{fig:figure_1_setup}. Although the photonic state evolves on the scale of hundreds of picoseconds, we achieve a time-resolved degree of entanglement up to a maximum 89$\%$ fully entangled fraction (FEF), opening in this way further scenarios for the exploitation of QD devices for quantum communication purposes in an out-of-the-lab framework.
\begin{figure*}
 \centering
    \includegraphics[width=\textwidth]{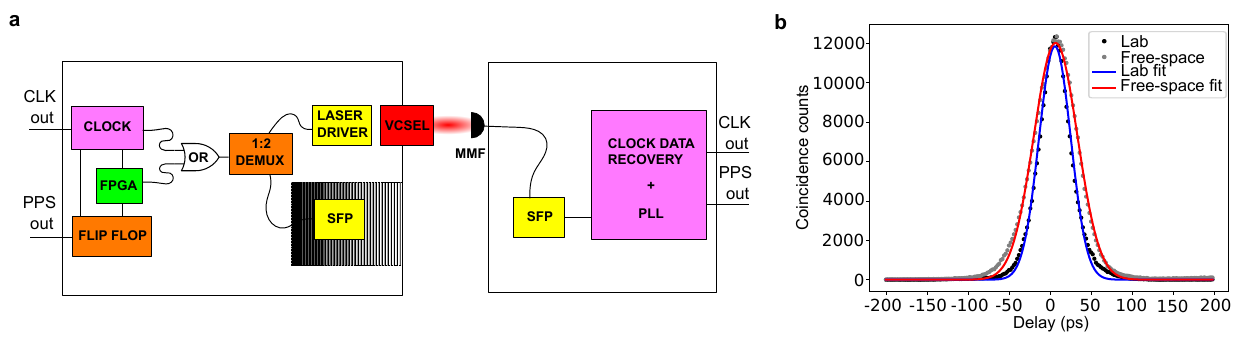}
    \caption{\textbf{Synchronization device:} \textit{\textbf{a} scheme of the synchronization device. A multiplexer sends the Clock and PPS signals to either an 852 nm VCSEL (free-space transmission) or a 1550 nm SFP module (fiber transmission), which we do not use in our demonstration, but would in principle allow three-node synchronization. The receiver unit is equipped with a SFP slot for optical input, converting the signal for a Clock and Data Recovery component with an integrated Phase-Locked Loop (PLL) to output clean clock and PPS signals. \textbf{b} comparison of time jitter of coincidence detection in the lab and over the free-space channel. We compute the time distribution of laser pulses traveling through the optical setup with respect to the pulse as immediately generated. First we consider the experimental setup in the laboratory (local case), using the same time correlator for the reference and propagated signal. Then, we measure the laser pulses in an equivalent setup at the end of the free-space channel, synchronizing two remote time correlators through the qssys device. By fitting the coincidence distributions with a Gaussian function, we find that synchronization across the free-space channel (red curve), using our parameters, only adds a time jitter of 39 ps FWHM with respect to an identical setup in the laboratory (blue curve).}}
    \label{fig:figure_2_sync_device}
\end{figure*}
\section{Results}
\begin{figure*}[!t]
    \centering
    \includegraphics[width=\textwidth]{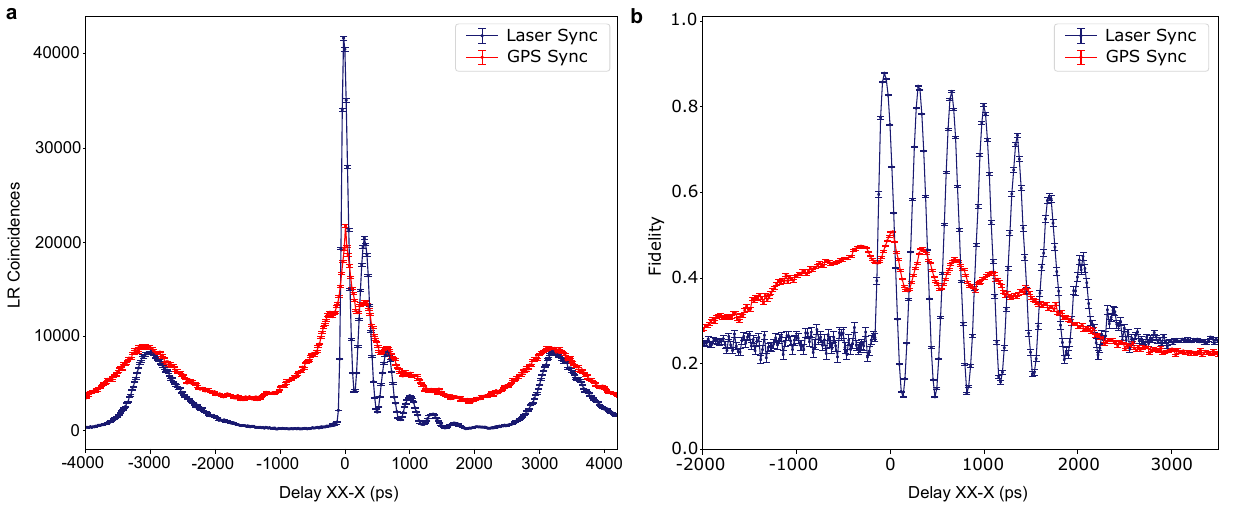}
    \caption{\textbf{Bell state oscillations over the free-space channel:} \textit{\textbf{a} population oscillations measured for the projector $\ket{LR}\bra{LR}$, employing the laser-based synchronization system or the GPS-disciplined oscillators, for a 30 minutes total integration time. \textbf{b}
     fidelity to the $\ket{\phi^+}$ state of the distributed two-photon state, employing the laser-based synchronization system or the GPS-disciplined oscillators. The laser synchronization allows to reliably distribute entanglement, enabling possible quantum communication application, by time-wise post-selection or specific quantum cryptographic protocol.}}
    \label{fig:figure_3_laser_vs_GPS_sync}
\end{figure*}

\begin{figure*}[t]
    \centering
    \includegraphics[width=\textwidth]{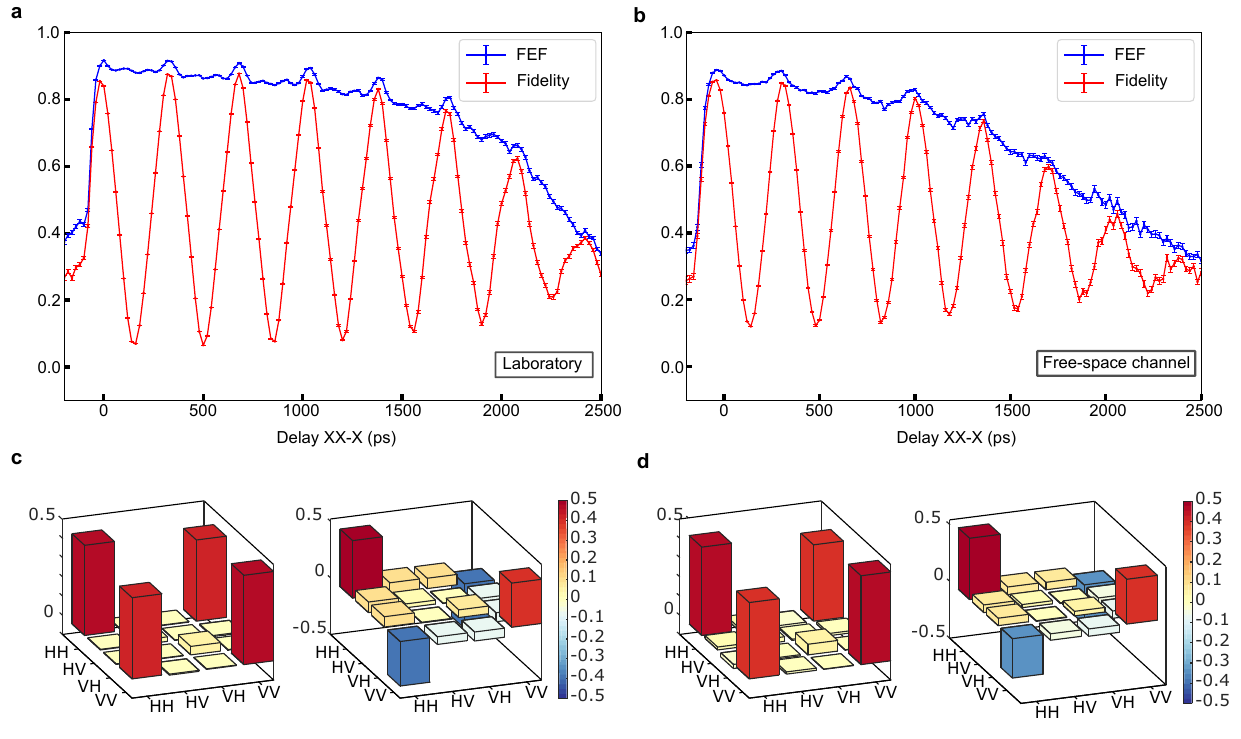}
    \caption{\textbf{Time-resolved entanglement distribution:}\textit{ fully entangled fraction (FEF) and fidelity to the $\ket{\phi^+}$state computed for XX-X coincidence events as a function of the detection delay as measured \textbf{a} in the laboratory and \textbf{b} across the free-space channel. Each considered time bin is 20ps wide for both experiments.
     The degree of entanglement remains high within the lifetime of the X transition, as expected, then it decays due to the decrease of the signal to noise ratio and the effect of nuclear spin noise \cite{schimpf2023hyperfine}.  
 Small oscillations of the FEF can be observed due to spurious radiative contributions which could not be separated by the XX-X-0 cascade signal. Error bars are computed by Montecarlo methods considering Poissonian statistics of detection events. The real part of the density matrices corresponding to two consecutive extremal points of the oscillations is reported for the \textbf{c} laboratory measurements, yielding $F_{\ket{\phi^+}}=87.7\pm0.1\%$ (left) and $F_{\ket{\phi^-}}=84.2\pm0.1\%$ (right), and for \textbf{d} the free-space channel experiment, yielding $F_{\ket{\phi^+}}=85.7\pm0.1\%$ (left) and $F_{\ket{\phi^-}}=81.3\pm0.1\%$ (right).}}
\label{fig:figure_4_entanglement_vs_delay}
\end{figure*}

As entangled photon emitter, we employ a GaAs QD generating photon pairs by the biexciton-exciton (XX-X) two-photon radiative cascade (inset of Fig. \ref{fig:figure_1_setup}), pumped by the Two-Photon Excitation (TPE) scheme \cite{brunner1994sharp,jayakumar2013deterministic}. The emission spectrum lies in the NIR wavelength range (around 785 nm), suitable for efficient free-space propagation \cite{abasifard2024ideal}. 
The XX-X cascade can be exploited for the on-demand generation of photon pairs with near-unity degree of entanglement \cite{huber2017highly}, ultra-low multi-photon emission probability \cite{hanschke2018quantum,neuwirth2022multipair} and high brightness, thanks to the use of photonic cavities \cite{liu2019solid,wang2019demand}. Nonetheless, QDs can be affected by geometrical asymmetries that induce a Fine Structure Splitting (FSS) between the two bright X states, resulting in an output two-photon polarization states that precedes with time \cite{gammon1996fine,seguin2005size}. 
Specifically, the emitted two-photon polarization state depends on the emission delay $\Delta\tau$ between the two photons, and can be written as:
\begin{equation}
    \ket{\phi}_{X,XX}=\frac{\ket{HH}+e^{i\frac{S}{\hbar}\Delta\tau}\ket{VV}}{\sqrt{2}}
    \label{eq:oscillating_state}
\end{equation}
where $S$ is the FSS value. From this description, we explicitly see that the two photons are maximally entangled for any $\Delta\tau$, but an indiscriminate integration over $\Delta \tau$ will lead to an effective trace operation over the temporal degree of freedom and the consequent loss of quantum correlation \cite{hudson2007coherence,fognini2019dephasing}.\\
FSS can be manipulated and tuned harnessing various external perturbations \cite{hudson2007coherence,vogel2007influence,gerardot2007manipulating,bennett2010electric,stevenson2006magnetic,seidl2006effect,plumhof2012experimental,kuklewicz2012electro}, and universally tuned with the use of multiaxial strain fields \cite{trotta2012universal,trotta2015energy,zhang2015high,wang2015towards}. 
However, the emitter we employ does not feature any of these FSS tuning mechanisms, but, in turn, it is equipped with a p-i-n diode structure, which allows strong suppression of blinking \cite{efros2016origin,paperQDdiode}. In addition, it is embedded in a circular Bragg resonator (CBR) \cite{rota2024source}, which enhances the collection efficiency up to $\eta_C\approx0.2$ without providing any significant Purcell acceleration, so that the XX lifetime amounts to $\tau_{XX}=194$ ps and the X lifetime  to $\tau_{X}=400$, likely due to the presence of the diode electric field \cite{undeutsch2025electric}.  However, these characteristics still allow us to pump the QD up to a 320MHz rate, multiplexing an originally 80 MHz pulse-shaped Ti:Sa laser thanks to a concatenation of Mach-Zehnder interferometers. We report the full characterization of the emitter in the Supplementary Material (SM).
The FSS of the chosen emitter is $S=11.5\pm0.5$ $\mu$eV, high enough to make Bell state oscillations clearly visible even in such a short emission time distribution. 
Specifically, we can estimate the oscillation period to be $T_S\approx360$ ps. Thanks to our fast detection system, composed of SNSPDs ($\approx$15 ps time jitter) and a high-resolution time correlator (2 ps time jitter), the experimental setup in our laboratory features an overall $21$ps FWHM time resolution for coincidence events, which is abundantly sufficient to reconstruct the oscillations of our QD two-photon entangled state.\\ 
As we step outside the laboratory, it is harder to maintain such low time jitter in coincidence measurements. The synchronization of the two parties at the ends of a free-space channel is usually performed by means of GPS-disciplined oscillators, characterized by a time jitter of around 400 ps \cite{steinlechner2017distribution,lu2022micius,bulla2023nonlocal,basso2021quantum,basset2023daylight,laneve2025quantum}. Thus, being the temporal uncertainty of coincidences detection comparable to the oscillations period, it would be impossible to clearly resolve the evolution of the remotely distributed two-photon state.
Therefore, we implement a novel synchronization approach, enabling us to greatly enhance the time resolution of joint photon detection across the free-space channel. We employ a prototype system realized by Quantum Space Systems GmbH (qssys), which modulates the 852 nm radiation of a diode laser with a clock and an reference (start) signal. The clock frequency can be tuned, and we chose a 200 MHz frequency for the optical signal, resulting in an electronical 4 MHz signal forwarded to the time correlators. The start signal can be set as pulse-per-second (PPS) but also to have a longer period. The period corresponding to our clock configuration was 2.5 s. These parameters were selected to ensure the best time jitter and stability throughout the experiment, as detailed in the SM. As depicted in Fig. \ref{fig:figure_1_setup}, the laser is sent from one of the two buildings of Sapienza's Physics Department (the Marconi building) to a second one (the Fermi building), traveling the free-space channel together with single photons. The laser signal is then spectrally separated from the photons, and part of it is employed to run stabilization systems that compensate for atmospheric turbulences via slow and fast steering mirrors \cite{basso2021quantum,basset2023daylight}. The remaining laser signal is coupled into a decoder device, which provides the time correlator located in the Fermi building with the synchronization signal. Meanwhile, the encoder feeds the time correlator in the Marconi building with an identical clock and PPS signal as the ones modulating the laser emission.
This encoding-decoding system is depicted in Fig. \ref{fig:figure_2_sync_device}\textbf{a} and detailed in the SM: the encoder unit features an oscillator chip providing the Clock signal and a field-programmable-gate-array (FPGA) extracting the PPS signal from the master clock. The selected Clock and PPS are directly fed to a time correlator, besides driving a VCSEL diode. The receiver unit converts the received laser and outputs clean clock and PPS signals to the time correlator at the other end of the free-space channel. This system achieves synchronization of the two remote detections sytems within an accuracy limited by a 39 ps jitter, as reported in Fig. \ref{fig:figure_2_sync_device}\textbf{b}, allowing to clearly resolve the Bell state oscillations of the QD two-photon emission, even when distributed across the free-space channel.
In Fig. \ref{fig:figure_3_laser_vs_GPS_sync}\textbf{a}, we report coincidence measurements sampled sending the X photons to the Fermi building through the free-space channel and measuring the XX photons in the Marconi building, projecting the polarization state in two orthogonal states of the circular basis. 
We explicitly compare measurements recorded employing the GPS synchronization and the laser-based synchronization system, showcasing the great advantage provided by the new system. 
The laser-driven synchronization clearly allows to fully resolve the coherent evolution of the entangled state, while the dynamics is not resolved by the GPS-based system. Moreover, the large time jitter of GPS synchronization causes the broadening of photon detection time distribution, which can lead to the overlap of events corresponding to different emission rounds. In this way, there is a non-negligible probability of assimilating correlated and uncorrelated distributed photon pairs, as evident in Fig. \ref{fig:figure_3_laser_vs_GPS_sync}\textbf{a}.
However, it can be noticed that the number of detected events is higher for the GPS-synchronized measurements. This is due to the fact that, in the current configuration, the GPS-driven system is more stable in time, granting a more efficient collection of coincidence events, thanks to the fact that the laser power can be entirely employed to feed the stabilization system. In the SM, we discuss this feature and possible improvements of the laser-based synchronization device. 
In conclusion, the oscillating entangled state can be distributed with a far higher fidelity to a target maximally entangled one using the laser-based system, as we show in Fig. \ref{fig:figure_3_laser_vs_GPS_sync}\textbf{b}, where we report the fidelity to the Bell state $\ket{\phi^+}=\frac{\ket{HH}+\ket{VV}}{\sqrt{2}}$:
\begin{equation}
\begin{split}
F_{\ket{\phi^+}}(\rho)=Tr(\rho \ket{\phi^+}\bra{\phi^+})=\\=\frac{1+<XX>-<YY>+<ZZ>}{4}
\end{split}
\end{equation}
for both synchronization methods, where X,Y,Z represent, respectively, the Pauli operators $\{\sigma_i\}_{i=1}^3$ in the polarization space and $\rho$ is the density matrix of the two photon polarization state. These measurements demonstrate that the new synchronization system allows the reliable distribution of a fast oscillating entangled state over the free-space channel.\\ 
Given these capabilities, we perform a time-resolved full quantum state tomography of the two-photon system distributed across the link, using the setup described in Fig. \ref{fig:figure_1_setup}. In this way, we reconstruct the time-resolved system density matrix, from which we can quantify the overall degree of entanglement between the two photons. We do this by computing figures of merit which are independent of 
the oscillating behavior expressed in \eqref{eq:oscillating_state} and are also independent of accidental polarization rotation induced by the setup.
 More specifically, from the density matrix it is possible to obtain the fully entangled fraction (FEF) \cite{grondalski2002fully}, representing the fidelity to the maximally entangled state that is closest to the real state of the system. This is an absolute measurement of the degree of entanglement, which also makes it possible to evaluate the actual suitability of the system for quantum information applications (such as QKD). 
To compute the density matrix, we perform 9 combinations of the three mutually unbiased polarization bases $\{H/V,D/A,R/L\}$, realizing an overcomplete 36 bases measurement \cite{altepeter2005photonic}. We integrate a total of 30 minutes for each combination, breaking the total measurement time in 1 minute measurements; this is necessary to ensure the stability of the synchronization system during each data take, as we explain in the SM. 
We reconstruct the density matrix using a maximum-likelihood estimation (MLE) algorithm \cite{vrehavcek2007diluted} for different time delays between X and XX detection events corresponding to the same generation, considering 20 ps wide time bins. Depending on the delay between the emission of the XX photon and the X photon, the joint polarization state will be different, yet maximally entangled in any case. 
From these data, we evaluate the degree of entanglement of the system depending on such emission delay.
We report in Fig. \ref{fig:figure_4_entanglement_vs_delay} the FEF and the fidelity to the $\ket{\phi^+}$ state computed as a function of the XX-X delay, comparing the values as measured in the lab (\textbf{a}) and across the free-space channel (\textbf{b}).
Examples of measured density matrices, both for the highest recorded $F_{\ket{\phi^+}}$ and $F_{\ket{\phi^-}}$ are reported in Fig. \ref{fig:figure_4_entanglement_vs_delay}\textbf{c}, for the laboratory experiment and \textbf{d} for the free-space channel distribution experiment.
Given its definition, we expect the FEF to remain constant in time, except for effects produced by noise or dephasing. 
Due to the  effective state mixing caused by the overall measurement time jitter, the maximum achievable FEF of the system distributed over the channel is lower than what is achieved in the lab, because of the lower time resolution; still, thanks to the improved synchronization accuracy, it is possible to retain up to FEF$=88.7\%\pm0.1\%$ over a maximum of $91.6\%\pm0.1\%$ measured in the lab. At the same time, the maximum fidelity we record across the channel is $85.7\pm0.1\%$ compared to the reference $87.7\pm0.1\%$ in the lab.  
A quantitative simulation study regarding the effect of detection time jitter on the achievable distributed degree of entanglement is reported in the SM. 
However, the entangled state we produce and distribute looks affected by some coherent noise contribution: this is evident from the oscillating behavior of the FEF. 
In the SM, we provide further analysis of the phenomenon, which we attribute to the presence of additional radiative processes, the contribution of which could not be filtered out.\\ 
The next question is whether the entanglement distribution we implement can be efficient enough to enable secure quantum communication. In particular, we consider explicitly Quantum Key Distribution (QKD) as a benchmark. We can evaluate the expected performance of the distributed state, post-selected bin-by-bin, and employed for an Ekert91 protocol \cite{ekert1991quantum} (or its advanced asymmetric version from Acin et al. \cite{acin2006bell}). In practice, this approach would be extremely inefficient, since the signal which does not provide a positive secure key rate would be discarded in the procedure. For example, if we only consider photon pairs in the time-bin yielding the highest FEF, we only account for less than the $0.01\%$ of the total QD emission. Therefore, we compute the secure key fraction (SKF) for every XX-X delay as a benchmark of the channel capability to resolve the evolving entangled state, rather than to suggest an actual application of our system.
For every time bin, we compute the QBER on the time-resolved density matrix as $QBER=\frac{1-E(Z,Z)}{2}$ and estimate the value of the CHSH violation $P$ optimized for the $\ket{\phi^+}$ state. Then, we compute the corresponding Devetak-Winter rate (for the non-device-independent case \cite{pironio2009device}):
\begin{equation}
r_{DW}=1-h(Q)-h\big(Q+\frac{P}{2\sqrt{2}}\big)
\end{equation}
where $h$ is the binary entropy. 
Over all the possible XX-X delays, we obtain a maximum time-resolved post-selected asymptotical $SKF=\max(0,r_{DW})=0.14\pm0.01$, which is compatible with secure key generation. However, we can also estimate the QBER in the case of perfect polarization compensation as $QBER=(1-FEF)/2$, since the FEF is robust to accidental rotations of the photon polarization. 
Thus, we consider the highest time-resolved $FEF=89\%$, which also corresponds to the highest time-resolved $P$, and we obtain both the best CHSH violation and QBER for the same XX-X delay, corresponding to a maximum $SKF=0.16\pm0.01$.
In conclusion, our novel synchronization system can support the exchange of secure key in a free-space channel even with a FSS value above 10 $\mu$eV, by implementing time-wise post-selection.\\ 
However, a more significant application of our results could be found in different protocols, such as the one proposed in \cite{pennacchietti2024oscillating}, \textit{ad hoc} designed to perform secure QKD employing oscillating Bell states.
The protocol theoretically devised in \cite{pennacchietti2024oscillating} is based on the six-state protocol from \cite{bruss1998optimal}, except for the fact that only events detected on the $\{H,V\}$ basis measurement are considered to write the secret key; in this way, it would be possible to employ even a QD affected by a large FSS to perform secure QKD. Since the three bases used for the protocol form a tomographically complete measurement, it is possible to use a random portion of the exchanged photons to compute the system's density matrix, so as to evaluate the FEF or concurrence \cite{wootters1998entanglement} and detect possible eavesdropping. It is worth noting that, for potential out-of-the-lab applications, this protocol would heavily rely on low time jitter between the remote parties, unlike the one which could be provided by GPS-based synchronization.
Thus, high time resolution is necessary to ensure eavesdropping detection by density matrix sampling, but it also delivers a further beneficial effect: as it can be deduced from Fig. \ref{fig:figure_3_laser_vs_GPS_sync}, a low coincidence detection jitter avoids the mixing of coincidences between photons generated by distinct excitation laser pulses. In this way, noise levels decrease, granting a better QBER, hence a higher SKF.
\section{Discussion}
We report the faithful distribution of an oscillating entangled state between the two ends of an urban free-space channel. To the best of our knowledge, our experiment integrates for the first time state-of-the-art time resolution in an entanglement distribution protocol conducted over a free-space quantum link.\\
The high-time resolution we achieve positively impacts the reliability of quantum correlation distribution in several ways. First of all, it allows accurate time-wise post-selection: even in our challenging case, where the oscillations period is comparable to the emitter excitation lifetimes, the fidelity to the ideal target state is enough to violate CHSH inequalities. Moreover, it allows to reduce the impact of noise originating from uncorrelated coincidence events, as showcased in Fig. \ref{fig:figure_3_laser_vs_GPS_sync}. This advantage can be applied to any high-rate source of entangled photons, both deterministic and non-deterministic, allowing to push the limit of entanglement distribution rates over free-space links.\\
Finally, our demonstration paves the way for direct applications of oscillating two-photon entangled state in quantum communication in realistic scenarios. Indeed, the ability to accurately synchronize remote detection sytems enables time-resolved QKD, both employing post-selection or suitable protocols \cite{pennacchietti2024oscillating}. 
Moreover, the possibility to have low time jitter in entanglement distribution is also significant beyond QKD, enabling spoofing-safe quantum clock synchronization based on entanglement \cite{jozsa2000quantum,giovannetti2001quantum}, or position determination \cite{giovannetti2002positioning}, for which oscillating states can be fruitfully exploited \cite{alqedra2026entanglement}. Our results provide a benchmark for the possibility to implement time-sensitive protocols in free-space networks, such as satellite-based quantum communication infrastructures.
In this framework, a bright deterministic source of entangled photons whose state evolves in time, as the one we employ, will be useful. 
It is worth mentioning that we do not address in any way the FSS of the QD: as discussed above, many methods have been developed to reduce FSS, even to erase it, but the most effective of them require a high level of technological sophistication, making the emitters' fabrication stage more challenging and less reliable. Because of that, efforts have been devoted to show that QD featuring high FSS have the potential to be employed for quantum communication purposes. 
In this work, we showcase the possibility to integrate these emitters in practical quantum networks, strongly reducing the trade-off between quality and quantity of the signal. 
This will facilitate the employment of deterministic entangled photon emitters in practical quantum networks, where the requirement of brightness to overcome unavoidable losses is paramount. 
\end{multicols}
\medskip
\textbf{Supporting Material}\par \noindent
Supporting Material is available 
from the author.

\medskip
\textbf{Acknowledgements}\par
\noindent 
The authors acknowledge funding from the European Commission by project QUID (Quantum Italy Deployment) funded in the Digital Europe Programme under the grant agreement No 101091408 and from Project ECS 0000024 Rome Technopole, – CUP B83C22002820006, NRP Mission 4 Component 2 Investment 1.5, funded by the European Union – NextGenerationEU. This project has received funding from the Austrian Science Fund FWF via the Research Group FG5 (10.55776/FG5) and from the cluster of excellence quantA [10.55776/COE1] as well as the EU HE EIC Pathfinder challenges action under grant agreement No. 101115575, from the QuantERA II program that has received funding from the European Union’s Horizon 2020 research and innovation program under EPIQUE Project Grant Agreement No. 101135288, and Grant Agreement No. 101017733 via the projects QD-E-QKD and MEEDGARD (FFG Grants No. 891366 and 906046). P.M. and H.W. acknowledge financial support from the German Federal Ministry of Research, Technology and Space (BMFTR) under funding number 16KIS1673.
The financial support by the Austrian Federal Ministry for Digital and Economic Affairs, the National Foundation for Research, Technology and Development and the Christian Doppler Research Association is gratefully acknowledged.

\medskip

\bibliographystyle{MSP}
\bibliography{biblio.bib}

\end{document}